# Randomness in Relational Quantum Mechanics

Gary Gordon

*Advantages of RQM*

There has been a lot of confusion over the years on the nature of the randomness exhibited in quantum mechanical phenomena. The relational interpretation of quantum mechanics (RQM) introduced by Carlo Rovelli (1) provides a new perspective on this and many other issues debated in connection with the applied QM theory. These changes result from a number of assumptions that depart from those commonly found in other QM interpretations. For example, RQM assumes that all objects in the universe, regardless of complexity, are quantum systems, and that the state of any quantum system is always defined in relation to another reference quantum system with which it has previously interacted. The quantum state of an object at an instant can vary in relation to its valid reference systems, suggesting that quantum states are not ontologically real. Furthermore, quantum states do not collapse as the result of a measurement, eliminating some past interpretation difficulties associated with the "measurement problem".

There is direct analogy here to the concept of velocity in classical mechanics. One never says that the velocity of an object has some unique value. It only makes sense to say that the velocity of an object has some value relative to some other reference object. In the present context, the strictly relational nature of object properties is assumed to apply to all fundamental particles such as electrons or quarks (or their fields), and to all other objects that are formed from these fundamental particles. This would include, for example, atomic nuclei, atoms, molecules, and arbitrarily formed combinations of atoms and molecules. What are usually termed macroscopic objects are thus really just extensive combinations of these simpler objects, and are deemed here to be no different in their strictly relational character.

In some interpretations of quantum mechanics, macroscopic objects are considered to be fundamentally different from quantum systems. As an assumed consequence, macroscopic objects, when they interact with a quantum system, can in some cases produce a change in the state of the quantum system as a result of the interaction, and in the process reveal specific value(s) of some of the quantum system's properties. In such cases, the macroscopic object is then said to have performed a "measurement" of the quantum system. In a typical such case, the measured quantum system is in a "mixed" state prior to such an interaction, consisting of the superposition of some number of "pure" states, where each of these pure states corresponds to a possible measurement outcome. Such mixed quantum states are defined statistically. That is, the specific pure state that results from a measurement cannot be predicted beforehand, but the relative probabilities of each of the possible outcomes is defined in the known initial state of the measured quantum system. The macroscopic system is said to "collapse" the initial mixed quantum state to the observed pure state corresponding to the measured variables.

The ability to produce a collapse of the state of a measured quantum system, producing deterministic results, is considered in such interpretations to be unique to macroscopic systems. This suggests that some criteria can be defined to operationally differentiate macroscopic and quantum systems. But the search for such criteria has been notoriously unsuccessful, a problem for such interpretations. Some interpretations address this issue by introducing a critical role for consciousness or intelligence in macroscopic systems capable of producing the collapse of a measured quantum state. But then do we suppose that the past operation of the laws of the universe underwent a huge change of some sort when its physical evolution initially produced such conscious or intelligent entities? In RQM, there is no need for such anthropologically motivated assumptions. State collapse issues are bypassed entirely in the "many world" QM interpretations, where it is assumed that each potential measurement outcome is actually realized in one of the multiple worlds that appear as a result of any measurement interaction. But, such solutions have their own problem of accepting an enormous proliferation of unseen worlds, and this has stood in the way of their wide acceptance.

Here, we maintain that RQM never has the need to invoke a state collapse mechanism in connection with a measurement process involving a second system. It was noted above that in RQM all systems are quantum in nature, although the additional complexity of typical macroscopic systems can render the evaluation of their quantum states from first principles a daunting or practically impossible task. (We return to this issue again below.) There are any number of cases where there is an initial randomness associated with the state of a quantum system before its interaction with a second reference quantum system. And after the interaction, the quantum state of the first system in relation to the second has specific values for the measured variables. Significantly, the cause of this loss of randomness is not the result of a state collapse, but simply the result of a change in the applied reference system to the one in the new interaction.

*Application to a dual-slit experiment*

To illustrate these assertions regarding state collapse, we consider a laser emitting the photons in a dual-slit interference experiment. The individual photons emerge from the laser at random points in time and randomly spread over a range of directions. This is often characterized by saying that each individual photon has a random quantum state, typically represented by a wavefunction. But if we adhere to RQM, the next question would be "state relative to what"? A relative state is defined when two quantum objects interact. In the interaction, some variables of each of the two objects become "known" to the other. Here, "known" does not imply a role for intelligence or consciousness. It simply refers to the fact that some properties of one of the interacting objects are revealed to the other as a result of the interaction. This result is reciprocal. Neither of two interacting quantum objects is the observer and the other the observed. They are each both observer and observed.

What does the laser learn about an individual photon that it emits? The likely answer is nothing! The laser has no practical way of knowing the emission time or direction of its individual emitted photons. And an emitted photon learns nothing relevant about the laser as a result of its emission. So, one might conclude that the individual photons emitted do not have a defined state during the interval between their emission by the laser and their impact at some point on either the slit screen, or (assuming that the photon passes through the slits) on the detection screen. This view can be debated perhaps, but there is no doubt that a new relational state of an emitted photon can be defined when it either impacts the slit screen or the detection screen. Then, a relational position state for the photon is defined at the point of impact. Of course, this change in the photon state is not a wavefunction collapse. At most it is a change in the state of the emitted photon from the (perhaps undefined) state relative to the laser, to the position state at the point of impact. In a similar way, state collapse need never be invoked in the RQM interpretation, avoiding many of the "measurement problems" associated in the past with other QM interpretations.

But we know that useful results (such as the distance between nulls in the resulting interference pattern on the detection screen) can be obtained by defining a wavefunction or other state representation for the photons randomly emitted by the laser. How is this done and what does it represent? To answer the second question first, the wavefunction actually represents properties of the statistical ensemble of photons, but not any of the individual photons. That reinforces the point that a quantum system wavefunction is epistemic in nature rather than ontic. In this case it represents statistical information about the ensemble of photons emitted in the experiment. This wavefunction is not usually computed from fundamental principles and the properties of the laser. The details of the wavefunction may not even be of interest. For example, the distance between the nulls in the interference pattern really only depends on the applicable geometries and the wavelength of the photons (assumed monochromatic for the sharpest interference patterns).

If an emitted photon wavefunction is desired for the laser, it is usually obtained by operating the laser for a long-enough period directed at a solid detection screen and keeping a record of the frequency of photon detections and the spatial distribution of the detection points on the screen. This "sample space" of detections represents an approximation to the unknown wavefunction for the statistical ensemble of emitted photons. Note that this process is considered to represent multiple identical single-photon detection experiments. The conditions of these experiments must be sufficiently controlled and kept constant, so that the results of the measured sample space statistics can be meaningfully applied to the subsequent slit-screen experiment.

So you see that the randomness in QM is a non-trivial issue, and is actually an important part of the debate regarding competing interpretations of QM. Rovelli's RQM interpretation is seen to have a number of advantages over other interpretations, including a less troublesome treatment of the measurements made on quantum systems.